# A Multiple Criteria Decision Analysis based Approach to Remove Uncertainty in SMP Models

**GOKUL YENDURI, THIPPA REDDY GADEKALLU***

***School of Information Technology and Engineering, Vellore Institute of Technology, Vellore, Tamilnadu 632014, India.***

Gokul.yenduri@vit.ac.in, Thippareddy.*g@vit.ac.in*

## Abstract

The fields of automation and robotics are constantly progressing with the help of artificial intelligence (AI). Advanced AI technologies are serving humankind in a number of ways, from healthcare to manufacturing. Advanced automated machines are quite expensive, but the end output is supposed to be of the highest possible quality. Depending on the agility of requirements, these automation technologies can change dramatically. The likelihood of making changes to automation software is extremely high, so it must be updated regularly. If maintainability is not taken into account, it will have an impact on the entire system and increase maintenance costs, which results in the wastage of organizational assets. Estimating the maintainability of any advanced automated machines is a must before launching them into the market. Once the product is launched, it is not feasible to estimate the change. Many companies use different programming paradigms in developing advanced automated machines based on client requirements. Therefore, it is essential to estimate the maintainability of heterogeneous software. As a result of the lack of widespread consensus on software maintainability prediction (SPM) methodologies, individuals and businesses are left perplexed when it comes to determining the appropriate model for estimating the maintainability of software, which serves as the inspiration for this research. Data required for the study was collected from the public repositories, which were used by several researchers in maintainability prediction. A structured methodology was designed, and the datasets were preprocessed and maintainability index (MI) range was also found for all the datasets expect for UIMS and QUES, the metric CHANGE is used for UIMS and QUES. The methods have been evaluated using metrics, RMSE, MAE, and $R^2$ over six popular techniques, namely, step-wise regression, support vector machine (SVM), neural networks (NN), multivariate adaptive regression splines (MARS), genetic algorithm optimized random forest (GARF), and classification and regression tree (CART). To remove the uncertainty among the aforementioned techniques, a popular multiple criteria decision-making model, namely the technique for order preference by similarity to ideal solution (TOPSIS), is used in this work. TOPSIS revealed that GARF outperforms the other considered techniques in predicting the maintainability of heterogeneous automated software.

Machine Learning, Maintainability, Code metrics, MCDA.

**Nomenclature**

| Acronym | Description |
|---|---|
| **AI** | Artificial Intelligence |
| **ANN** | Artificial neural networks |
| **BM** | Base Model |
| **C&K** | Chidamber & Kemerer |
| **CART** | Classification And Regression Tree |
| **CKJM** | Chidamber and Kemerer Java Metrics |
| **CPMP** | Cross-Project maintainability Prediction |
| **FLANN** | Functional Link Artificial Neural Network |
| **GARF** | Genetic Algorithm Optimized Random Forest |
| **LOC** | Line of Code |
| **MAE** | Mean absolute error |
| **MARS** | Multivariate adaptive regression splines |
| **MCDA** | Multiple criteria decision analysis |
| **MI** | Maintainability index |
| **NB** | Naive Bayes |
| **NN** | Neural Network |
| **O and H** | Oman and Hagemeister model |
| **OO** | Object-Oriented |
| **PCC** | Pearson correlation coefficient |
| **$R^2$** | R square |
| **RMSE** | Random Mean Square Error |
| **SLOC** | Source Line of Code |
| **SMP** | Software maintainability prediction |
| **SVM** | Support vector machine |
| **SWR** | Step wise regression |

## 1. INTRODUCTION:

In today's world, innovation is taking place rapidly, and as a result, the world is moving toward complete automation with the help of AI. The combination of AI and automation is very fascinating and covers so many problems, like deliberate action, planning, understanding open environments, and interacting with other robots [1]. The AI and automation duo have had a lot of impact on many fields, like employment, processing mathematical problems, the hospitality industry, smart cities, and many others. Robotics has always proved to be productive for AI research. This combination also increases the capabilities of robotics in assisting humans at factories, hospitals, farms, etc. [2]. As a result, there has been a significant increase in the robotic software sector, and clients expect high-quality software, which has become a challenge for software developers [3]. Companies are competing to bring the best-automated products to market and provide a large number of upgrades to already released products to stay ahead in this competitive market [4].

According to ISO/IEC 25010 [5], maintainability is a critical factor for product quality. As the automated product's software is highly agile, it is always evolving to meet the needs of the clients. So, the maintainability of a product should be given top priority before it is released to the market as it directly affects the cost. The distinguishing characteristic of any software is "change," and as a result, extra care should be taken while developing software. As a result, the software system is expected to be easily modified. The maintainability of a product's software is reliant on internal and external quality factors. Researchers have estimated software maintainability using the MI, structural measures, code smells, and size [6]. The MI is most admired among the researchers as it is validated by Hewlett-Packard.

The MI in Eq. (1) was proposed by Coleman et al. (1994) [7]. It is a combination of several metrics, including Halstead's Volume (HV), McCabe's cyclomatic complexity (CC), lines of code (LOC), and percentage of comments (COM), and is validated by HP [3].

$$MI = 171 - 5.2lnV - 0.23G - 16.2lnL \qquad (1)$$

The Software Engineering Institute (SEI) has derived an MI in Eq. (2), which is based on the Coleman work in the year 1997. The MI ranges from 0–100, where 0 indicates that the software is hard to maintain and as the range increases towards 100, it indicates that the software is maintainable [8].

$$MI = 171 - 5.2log_2V - 0.23G - 16.2log_2L + 50sin(\sqrt{2.4C}) \qquad (2)$$

Radon [9], a popular Python tool, uses another derivative, MI, as depicted in Eq. (3). It is calculated based on the following SEI and Visual Studio derivatives.

$$MI = max[0,100171 - 5.2lnV - 0.23G - 16.2lnL + 50sin50sin(\sqrt{2.4C}))](3)$$

In the year 2011, Microsoft team blog [10] has reset the ranges or thresholds to 0-9 = Red, 10-19 = Yellow and 20-100 = Green which indicates poor medium, and high range of maintainability with minor modifications as shown below in Eq. (4).

$$MI = \frac{MAX(0, (171 - 5.2 * ln(HV) - 0.23 * (CC) - 16.2 * ln(LOC)) * 100)}{171} \qquad (4)$$

Though MI has attracted a lot of attention for its widespread use, there is some criticism for its use of fewer metrics and its compatibility with new programming paradigms. The uncertainty in predicting software maintainability in heterogeneous software and the use of SMP techniques remains difficult to overcome. Our research focuses on looking at how well popular software maintainability prediction models (SMP) work on different types of datasets. This helps get rid of the large amount of uncertainty that comes with choosing the appropriate prediction model.

The following are the proposed method's unique contributions:

1. Applying well-known SMP techniques to a heterogeneous dataset and determining which SPM technique is best for each dataset.
2. This is a novel contribution in which MCDA is used to evaluate the performance of SMP techniques, thereby assisting in the removal of model ambiguity.

The rest of the paper is structured as follows. Recent works are discussed in Section 2. The proposed methodology for predicting software maintainability is detailed in Section 3. Section 4 describes the techniques that are used, along with the results and discussion. Section 5 discusses potential threats to the proposed work's validity. Section 6 includes the research findings and detailed conclusions.

## 2. Recent Works

Software maintainability is hard to measure directly. So, previous researchers used predictive models. Measuring the software's maintainability is dependent on the metrics used, and identifying the appropriate metrics for source code is the most difficult aspect of maintainability prediction. However, no commonly accepted strategy for identifying the relevant metrics or prediction models is available [11]. Researchers have employed statistical approaches, machine learning algorithms, deep learning algorithms [12], and ensemble models. There is a scarcity of the data that is needed to analyze cross-domain projects and other benchmark datasets.

In 2021, Iqbal et al. [13] used a supervised learning approach to identify the changes that were required in the legacy system's current software components. New requirements and defect kinds necessitated a thorough redesign of the software components' interfaces. The software maintainability was assessed using the naive bayes classifier, a machine learning technique. Software components designed with the inverse criteria in mind were found to be error-free and easily adaptable to client needs. The authors employed limited datasets to estimate maintainability using only one machine learning technique, which is a significant flaw in this work. It can be encapsulated using heterogeneous software and recent algorithms.

In 2021, Lakra et al. [14] applied hyperparameter tuning on five regression–based ML algorithms like random forest, ridge regression, support vector regression, stochastic gradient descent, and gaussian process regression for two commercial object-oriented datasets, namely QUES and UIMS. The results exhibited substantial improvements when compared to the existing base models. The work primarily focuses only on fine-tuning models, lacks the usage of heterogeneous software, and has the limitation of not addressing the uncertainty in software maintainability prediction models. The work also uses only a few datasets, which is a serious drawback.

In 2020, Elmidaoui et al. [15] conducted a study on empirical evidence for the accuracy of software product maintainability prediction (SPMP) using ML techniques. The after-effects of about 77 studies that were published between 2000 and 2018 are inspected in this study based on the following criteria: maintainability prediction approaches, validation methods, accuracy criteria, the overall accuracy of ML techniques, and the techniques with the best performance. In the maximum number of studies, ML techniques' performance exceeded the non-ML techniques', such as regression analysis (RA), whereas fuzzy and neuro-fuzzy (FNF) outscored SVM/R, DT, and ANN. When several techniques were claimed to be superior, no specific technique could be recognized as the best, which is a serious limitation in this work.

In 2020, Malhotra et al. [16] used nine oversampling and three under-sampling approaches on unbalanced data. A detailed comparison of fourteen ML and fourteen search-based strategies has been taken into consideration to predict the class maintainability. This work supports the use of the Safe-Level Synthetic Minority Oversampling Technique (Safe-SMOTE) in handling imbalanced data when predicting class maintainability. This work has certain limitations compared to heterogeneous techniques and recent ML techniques. This work has an ambiguity in choosing an appropriate algorithm for estimating maintainability, which is a major drawback.

In 2020, Shikha Gupta et al. [17] proposed an enhanced-RFA (Random Forest Algorithm) technique for software maintainability prediction. The suggested method combines the random forest (RF) algorithm with three widely used feature selection techniques: chi-square, RF, and linear correlation filter, as well as a re-sampling strategy to increase the core RF algorithm's prediction accuracy. Using $R^2$, the performance of enhanced-RFA is assessed on two commercially available datasets, namely QUES and UIMS. The proposed approach performs much better than RFA for the specified datasets using chi-square, RF, and linear correlation filter approaches. This work has limitations in comparison to heterogeneous techniques and does not compare itself to recent ML techniques.

In 2020, Malhotra et al. [18] implemented several ML, statistical (ST) and hybridization (HB) techniques to create prediction models for software maintainability in this work. The important finding is that ML-based models outperform ST models in terms of overall performance. The use of HB methods for software maintainability prediction is restricted. It is encouraging that this work has reported the prediction performance of a few models developed using HB techniques, but no conclusive results about the performance of any of these techniques are reported and this paper ignores the metaheuristics.

The summary of the recent works is depicted in Table 1.

| References | Year | Models Used | Best Models | Limitations |
|---|---|---|---|---|
| 13 | 2021 | Machine learning technique | Naive Bayes classifier | Limited datasets |
| 14 | 2021 | Machine learning technique with hyper parameter tuning | Grid search method | Uncertainty in software maintainability prediction models |
| 15 | 2020 | Machine learning techniques and statistical techniques | ML techniques | Uncertainty in software maintainability prediction models |
| 16 | 2020 | Machine learning techniques and search based (SB) techniques | Safe-SMOTE with ML technique | Certain limitations over heterogeneous techniques and recent ML techniques |
| 17 | 2020 | RF algorithm with re-sampling technique | RF algorithm with re-sampling technique | Limits over heterogeneous techniques and did not compare itself with recent ML techniques |
| 18 | 2020 | Machine learning techniques, Statistical (ST), and Hybridized (HB) techniques | Machine learning (ML) | No conclusive results about the performance of any of these techniques is reported |

**Table 1** Recent works

## 3.Predicting software maintainability

### 3.1 Experimental setup

The software maintainability prediction modes are implemented using R Programming. The aim is to reduce the prediction error and improve the robustness of the proposed model. GARF compares its performance over other models like step-wise regression, support vector machine, NN, MARS, and CART. The datasets are obtained from the LI & Henry [19] and PROMISE [24] repositories. The details of these datasets are described in Table 2:

.

| DATASET | LANGUAGE | FEATURE | INSTANCES | SOURCE |
|---|---|---|---|---|
| **UIMS** | JAVA | 11 | 39 | LI &HENRY [19,20,21,22,23] |
| **QUES** | JAVA | 11 | 71 | LI &HENRY [19,20,21,22,23] |
| **CM1** | C | 40 | 505 | PROMISE [24] |
| **JM1** | C | 21 | 10878 | PROMISE [24] |
| **KC1** | C++ | 21 | 2107 | PROMISE [24] |
| **KC3** | JAVA | 40 | 458 | PROMISE [24] |
| **MC1** | C&C++ | 39 | 9466 | PROMISE [24] |
| **MC2** | C | 40 | 161 | PROMISE [24] |
| **MW1** | C | 40 | 403 | PROMISE [24] |
| **PC1** | C | 40 | 1107 | PROMISE [24] |
| **PC2** | C | 40 | 5589 | PROMISE [24] |
| **PC3** | C | 40 | 1563 | PROMISE [24] |
| **PC4** | C | 40 | 1458 | PROMISE [24] |
| **PC5** | C++ | 39 | 17186 | PROMISE [24] |

**Table 2** Datasets used in this research

In contrast, the distinction is drawn in correspondence with the comparative analysis $R^2$, MAE and RMSE as shown in Fig. (2).

### 3.2 Objective Function

The motivation of this paper is to find the better software maintainability prediction model with less error ratio when applied to heterogeneous datasets. The error is estimated based on the difference between $MI^{actual}$ and the predicted $MI^{predicted}$, which is defined in Eq. (5).

$$error = \min\left(MI^{actual} - MI^{predicted}\right) \tag{5}$$

### 3.3 Solution Encoding

$\overline{W}$ that represents the optimized weight, and the solution proposed is illustrated in Fig. (1).

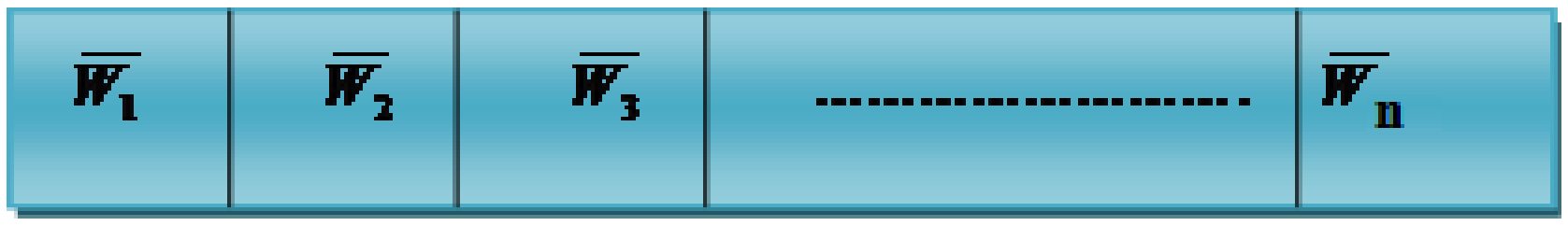

**Fig. 1.**Solution encoding

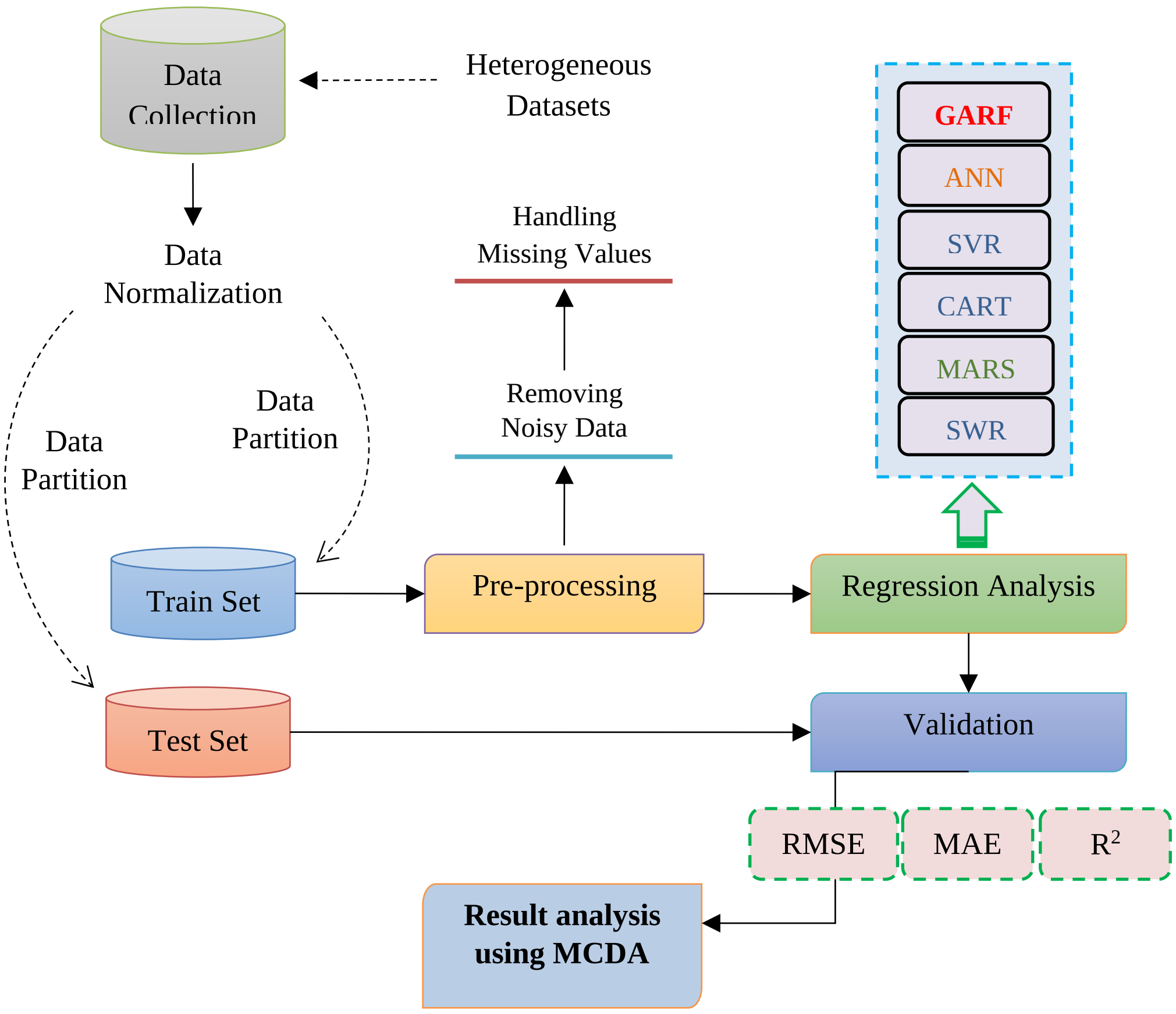

**Fig.2.** A methodology for assessing the maintainability of automated software

### 3.4 Genetic Algorithm

To improve prediction, this paper uses an evolutionary computing and most admired algorithm, GARF [25]. GARF algorithm imitates human evolution, in particular, gene evolution, and is inspired by charles darwin's theory of natural evolution. This algorithm represents the natural selection mechanism where the fittest individuals are chosen for succession to generate next-generation offspring. Parallelism is supported by the genetic algorithm, which is easily modified and adaptable to various problems. It is easy to disseminate and can search a large and diverse solution space. A non-knowledge-based optimization process is used to evaluate the fitness function. Finding the global optimum and avoiding becoming trapped in the local optimum is simple. A suite of potential solutions can be returned by multi-objective optimization. GARF is appropriate for large-scale and diverse optimization problems. The five rules of applying genetic algorithm are shown in Fig 3.

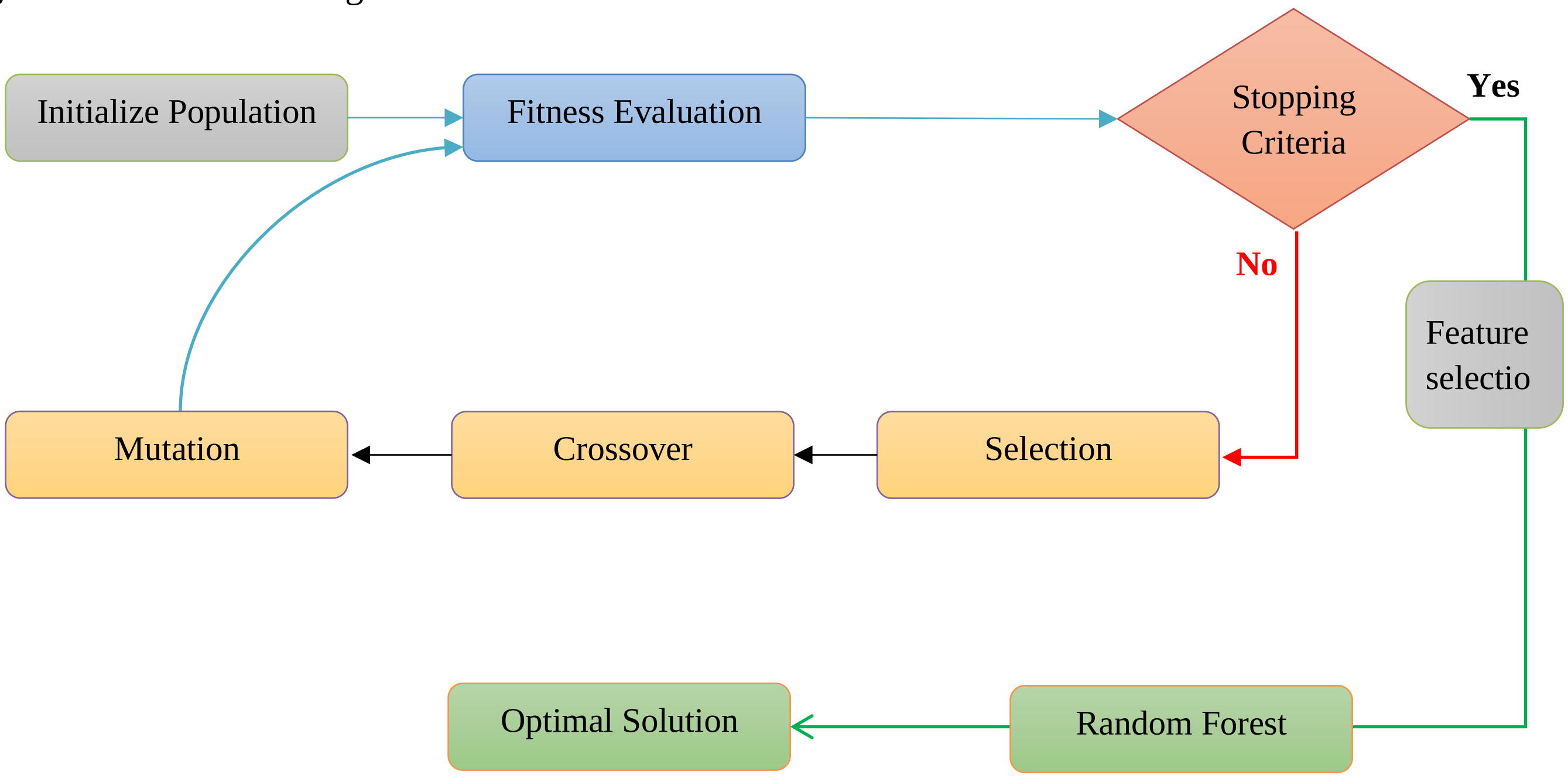

**Fig.3.** Genetic Algorithm Optimized Random Forest Algorithm

| **Algorithm 1: Pseudo Code of Genetic Algorithm for Feature Selection** | | |
|---|---|---|
| Commence | | |
| | Initialize time T: = 0; | |
| | Initialize the population W; | |
| | Evaluate Fitness of initial population E(W)=N; | |
| | Test for stopping criterion (time, fitness); | |
| | stopping criterion is YES (Optimal Solution OS(N)); | |
| | While (NO) | |
| | | Increment time T: = T+1; |

| | | | |
|---|---|---|---|
| | | | select a sub-population for offspring production $W^n$; |
| | | | Crossover the "genes" of selected parents C ($W^n$) = S; |
| | | | Mutate M(S)=P |
| | | | Evaluate Fitness E(P) =F |
| | | | Test for OS(F) |
| | | End | |
| | End | | |
| | Identify the features | | |
| | End while | | |
| End | | | |

While comparing GARF with other popular algorithms, GARF can deal with multi-model problems more efficiently. Further, it uses a good initial solution to start its iteration process. Simultaneously, the fitness of the whole population is evaluated and multiple individuals are stochastically selected from the current based on the fitness value then the new population is formed where the process iterates until the best solution is reached.

## 4. Results and Discussion

The proposed methodology for software maintainability prediction is implemented using R programming. The developed model aims to reduce the error rate. Six popular techniques, namely, SWR, SVM, NN, MARS, GARF, and CART are considered for software maintainability prediction and their performance is evaluated based on RMSE, MAE, and $R^2$. The computational complexity of GARF is 1200 seconds per iteration on average over considered datasets in a system i5 processer and 16 GB RAM.

### 4.1 Performance Analysis

A critical step in any empirical study is determining the predicted model's accuracy. The model predicts the value of the dependent variable, which is then compared to the actual value to discover errors. The current work compares various popular ML techniques, statistical (ST), and metaheuristic techniques using the following measures.

$$\boldsymbol{MAE} = 1/N \sum_{i=1}^{N} \frac{Actual\ value - Anticipated\ value}{Actual\ value} \quad (6)$$

The mean absolute error (MAE) [8] in Eq. (6), is a standardized measure used to find differences between the actual and anticipated values of a dependent variable. MAE calculates the difference between the actual and anticipated values first and then divides the result by the actual value. After that, each data point's absolute value is added together and divided by the entire number of data points.

$$\boldsymbol{RMSE} = (\sqrt{1/n\sum(Actual\ value - Predicted\ value)2} \tag{7}$$

The difference between predicted and actual values for each class is squared, then averaged, and finally the square root of the average value is calculated in RMSE [8] as Eq. (7).

In a regression model, $R^2$ [8] Eq. (8), is the amount of variation explained by an independent variable or factors for a dependent variable. The $R^2$ value indicates how much the variance of one variable explains the variance of the other.

$$\boldsymbol{R\ squared} = 1 - \frac{\text{First sum of errors}}{\text{Second sum of errors}} \tag{8}$$

In regression analysis, the residual squared error ($R^2$), mean absolute error (MAE), and root mean squared error (RMSE) metrics are used to assess the model's performance. The lower the value of MAE, RSQ, and RMSE, the more accurate a regression model is considered to be. Compared to the MSE, the mean squared error (MSE) and RMSE penalize the big prediction errors (MAE). However, because it has the same units as the dependent variable, RMSE is more commonly used to evaluate the efficiency of the regression model when compared to other random models than MSE (Y-axis).

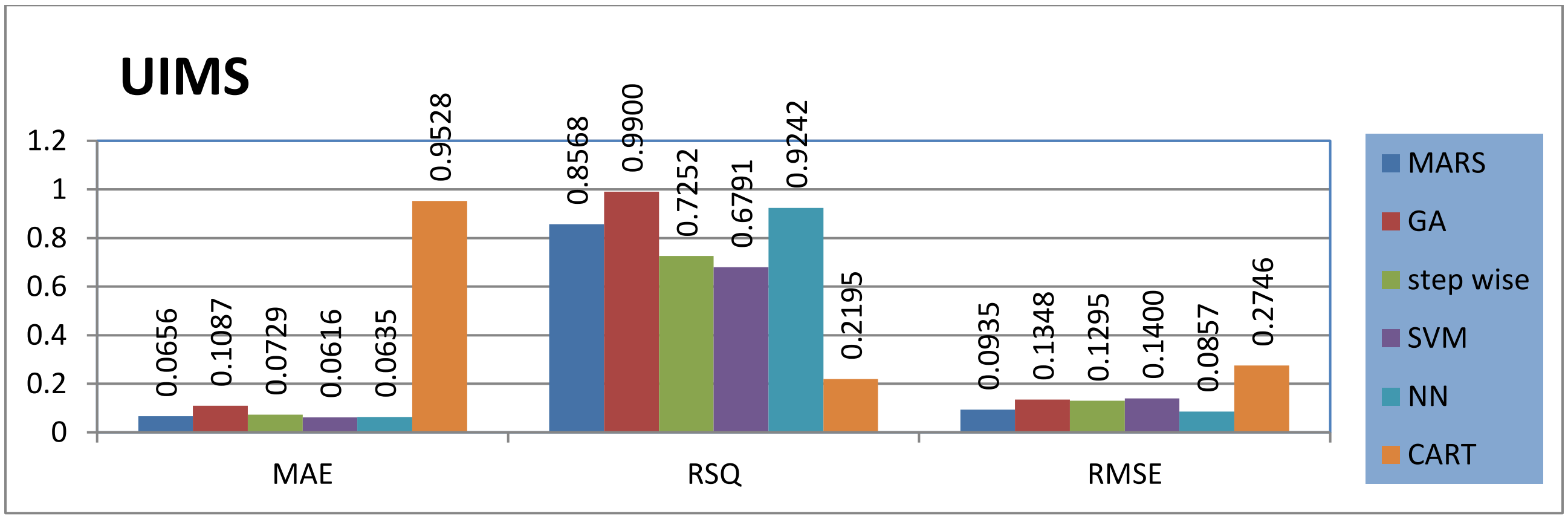

**Fig.4.** Analysis on renowned maintainability prediction models using evaluation measures on UIMS dataset.

The UIMS dataset is private and the source code is entirely developed using Java programming which contains 39 instances and 11 features. The genetic algorithm has achieved better performance compared to other models.

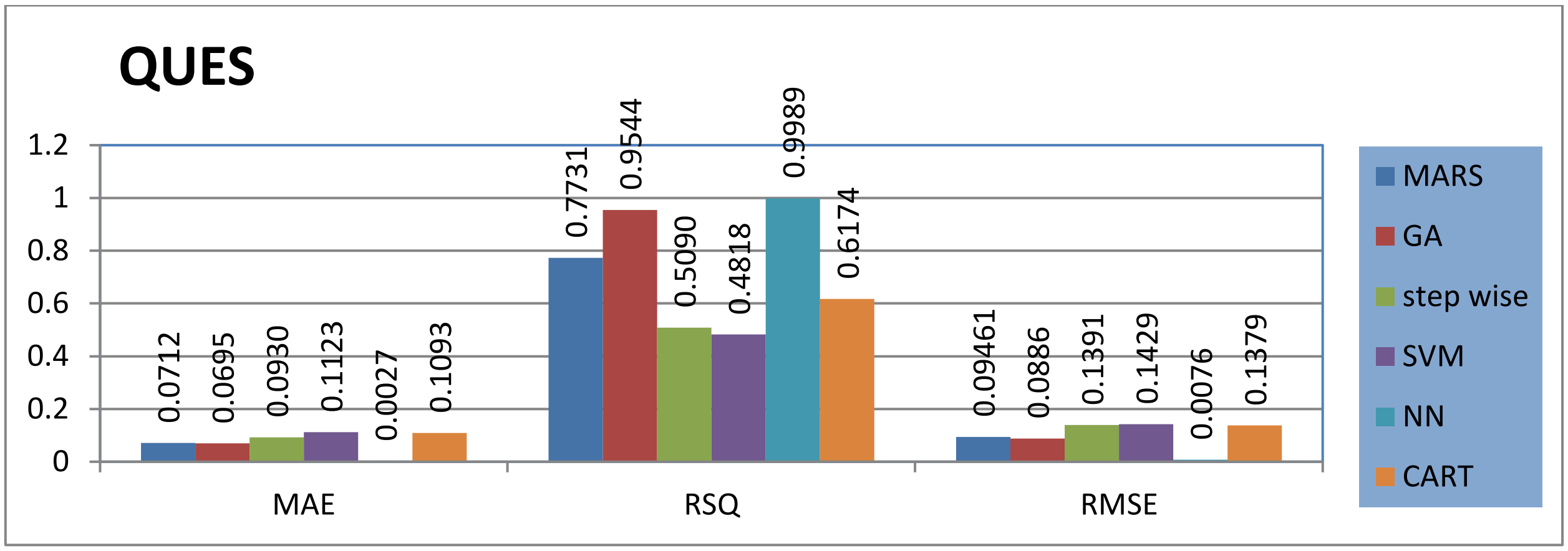

**Fig.5.** Analysis on renowned maintainability prediction models using evaluation measures on QUES dataset.

The QUES dataset is private and the source code is entirely developed using Java programming which contains 71 instances and 11 features. Neural networks have achieved better performance compared to other models.

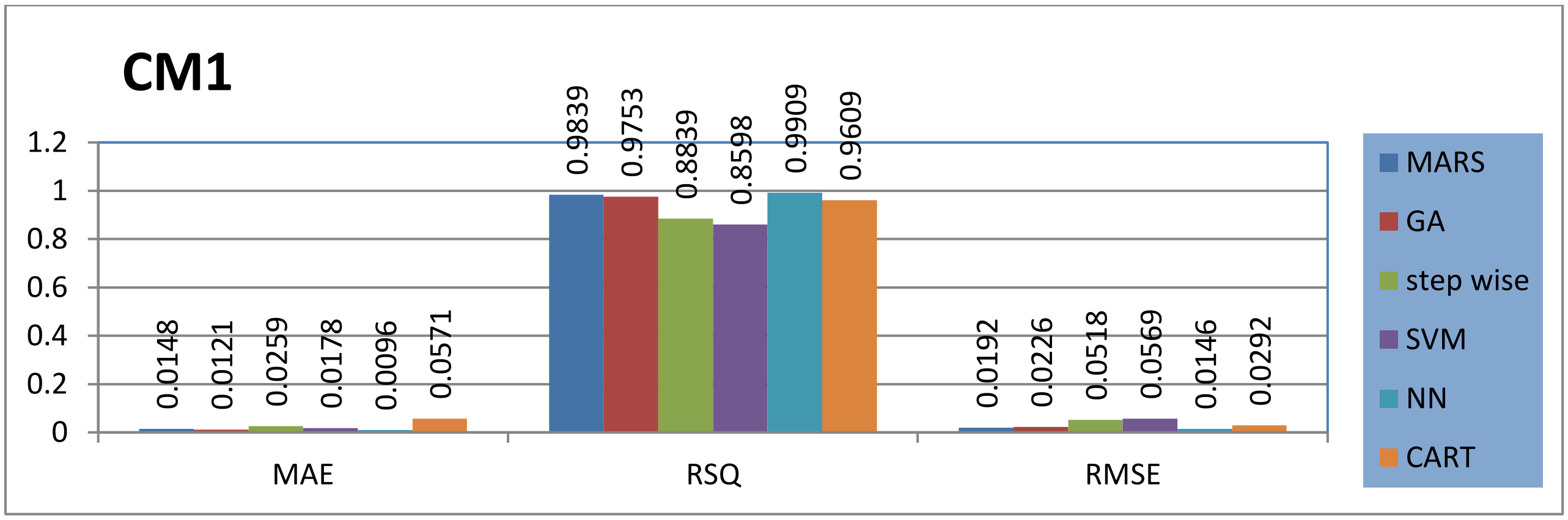

**Fig.6.**Analysis on renowned maintainability prediction models using evaluation measures on CM1 dataset.

The CM1 dataset is public and the source code is entirely developed using C programming which contains 505 instances and 40 features. Neural networks have achieved better performance compared to other models.

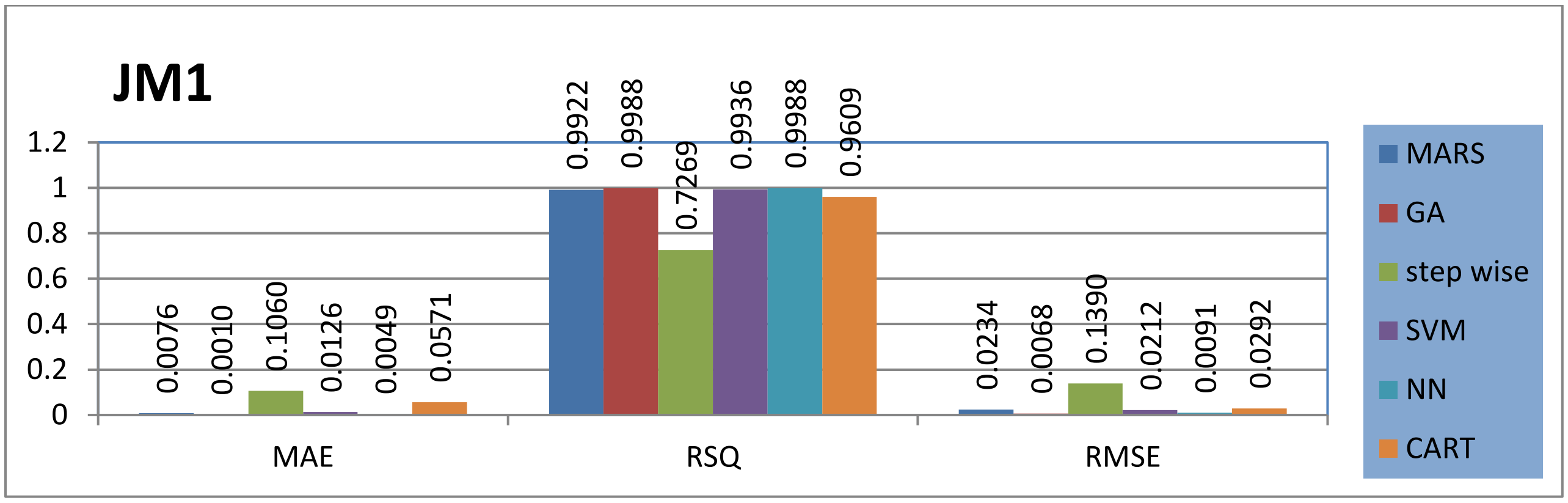

**Fig.7.** Analysis on renowned maintainability prediction models using evaluation measures on JM1 dataset.

The JM1 dataset is public and the source code is entirely developed using C programming which contains 10878 instances and 21 features. Neural networks [26] have achieved better performance compared to other models.

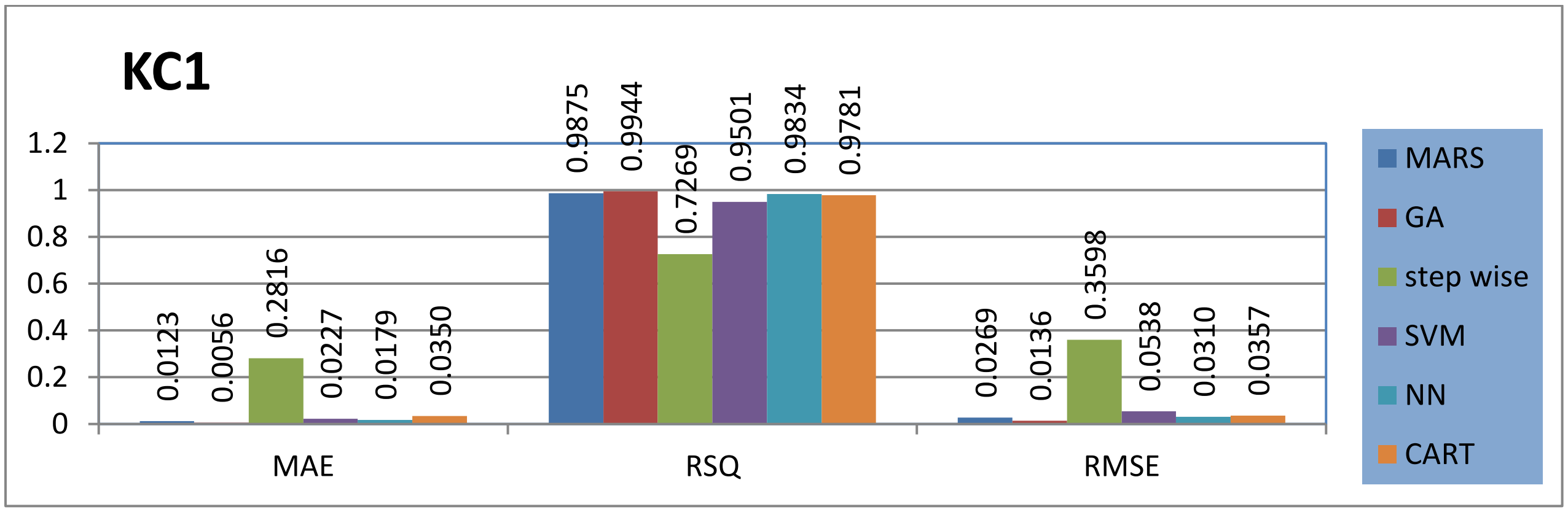

**Fig.8.** Analysis on renowned maintainability prediction models using evaluation measures on KC1 dataset.

The KC1 dataset is public and the source code is entirely developed using C++ programming which contains 2107 instances and 21 features. Genetic algorithm has achieved better performance compared to other models.

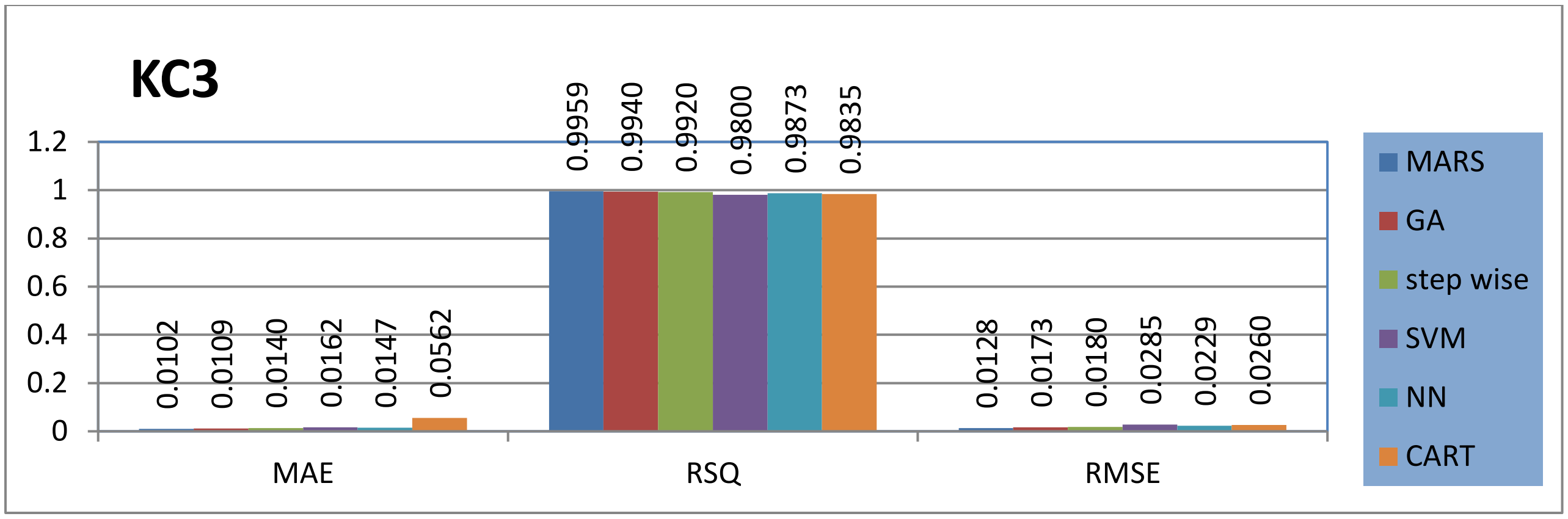

**Fig.9.** Analysis on renowned maintainability prediction models using evaluation measures on KC3dataset.

The KC3 dataset public and the source code is entirely developed using Java programming which contains 458 instances and 40 features. MARS have achieved better performance compared to other models.

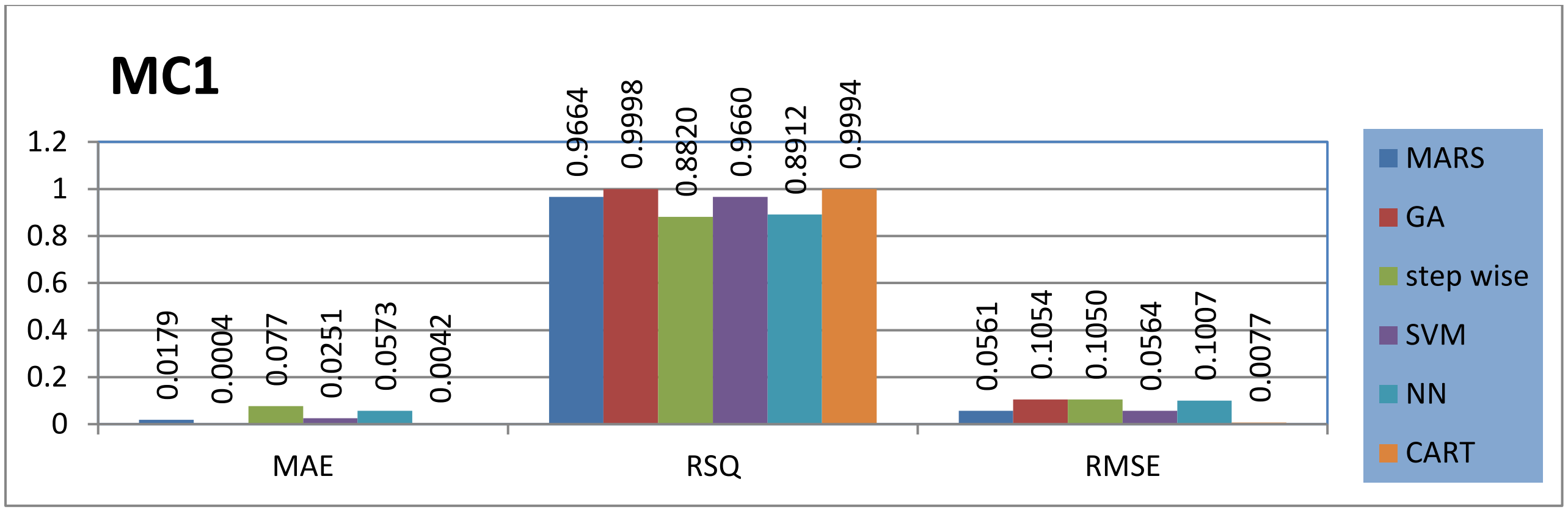

**Fig.10.** Analysis on renowned maintainability prediction models using evaluation measures on MC1 dataset.

The MC1 dataset is public and the source code is entirely developed using C and C++ programming which contains 9466 instances and 39 features. Genetic algorithm has achieved better performance compared to other models.

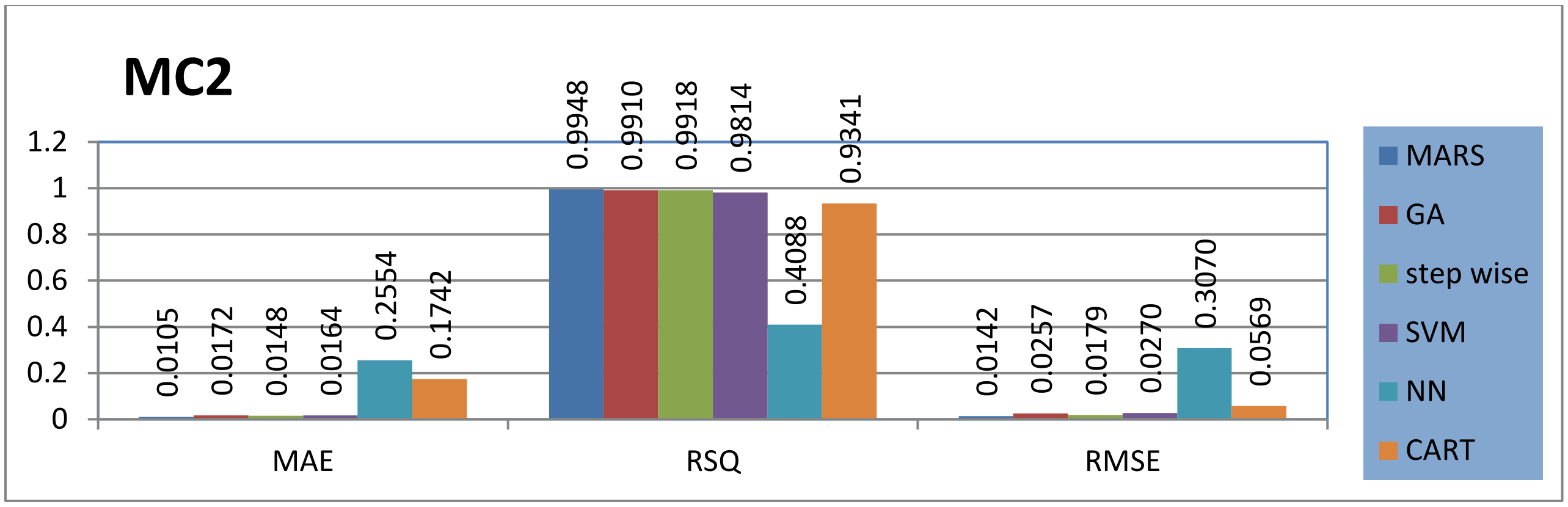

**Fig.11.** Analysis on renowned maintainability prediction models using evaluation measures on MC2 dataset.

The MC2 dataset is public and the source code is entirely developed using C programming which contains 161 instances and 40 features. MARS has achieved better performance compared to other models.

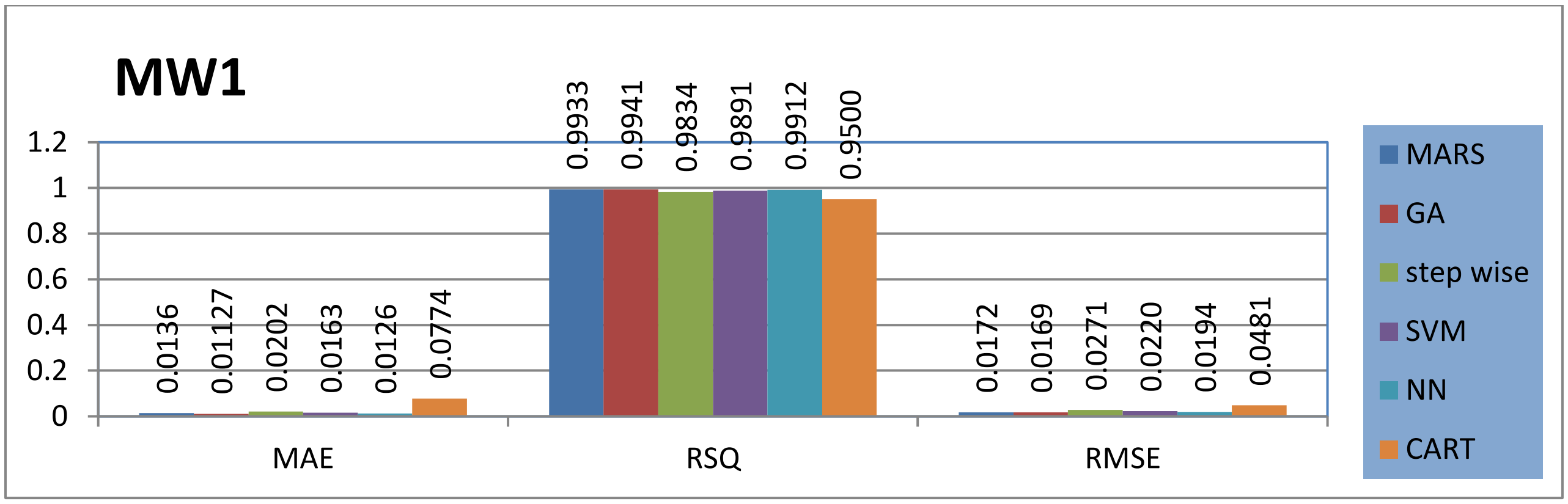

**Fig.12.** Analysis on renowned maintainability prediction models using evaluation measures on MW1 dataset.

The MW1 dataset is public and the source code is entirely developed using C programming which contains 403 instances and 40 features. Genetic algorithm has achieved better performance compared to other models.

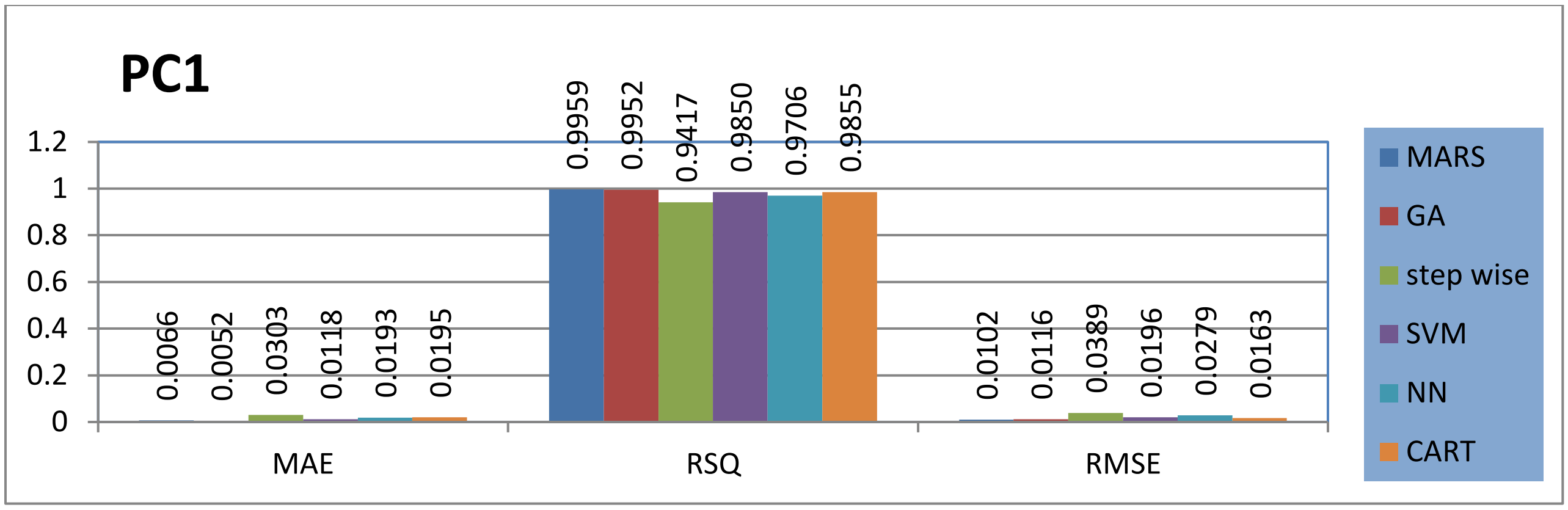

**Fig.13.** Analysis on renowned maintainability prediction models using evaluation measures on PC1 dataset.

The PC1 dataset is public and the source code is entirely developed using C programming which contains 1107 instances and 40 features. MARS has achieved better performance compared to other models.

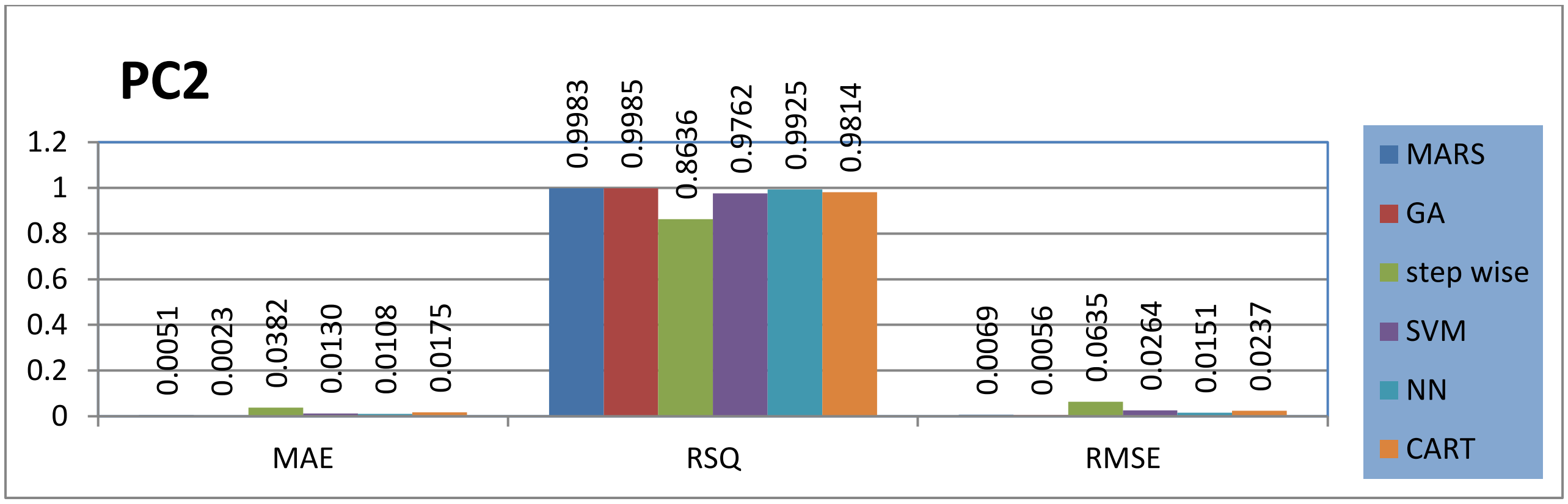

**Fig.14.** Analysis on renowned maintainability prediction models using evaluation measures on PC2 dataset.

The PC2 dataset is public and the source code is entirely developed using C programming which contains 5589 instances and 40 features. Genetic algorithm has achieved better performance compared to other models.

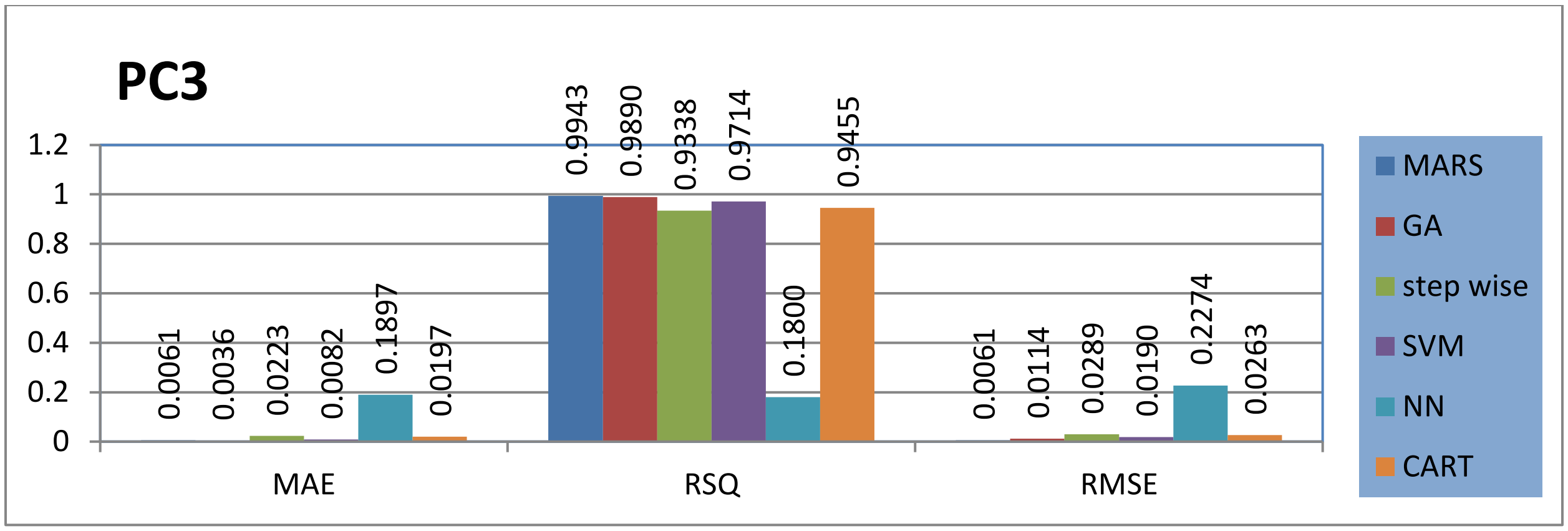

**Fig.15.** Analysis on renowned maintainability prediction models using evaluation measures on PC3 dataset.

The PC3 dataset is public and the source code is entirely developed using C programming which contains 1563 instances and 40 features. MARS has achieved better performance compared to other models.

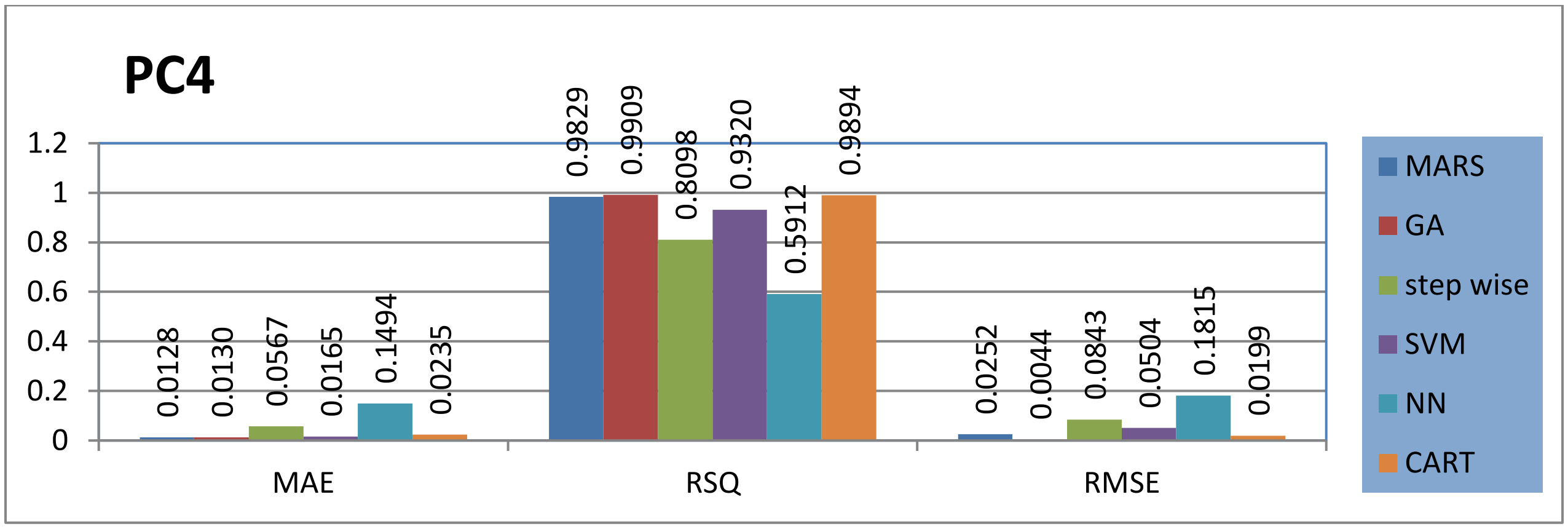

**Fig.16.** Analysis on renowned maintainability prediction models using evaluation measures on PC4 dataset.

The PC3 dataset is public and the source code is entirely developed using C programming which contains 1458 instances and 40 features. Genetic algorithm has achieved better performance compared to other models.

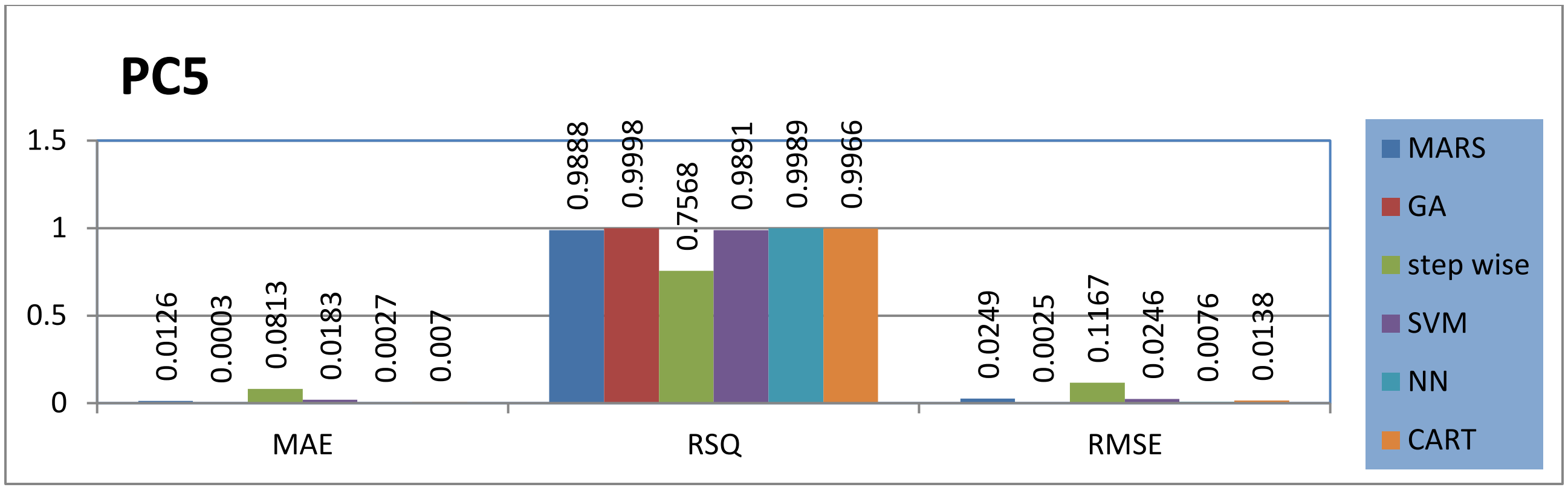

**Fig.17.** Analysis on renowned maintainability prediction models using evaluation measures on PC5 dataset.

The PC5 dataset is public and the source code is entirely developed using C programming which contains 17186 instances and 39 features. Genetic algorithm has achieved better performance compared to other models.

### 4.2 Technique for order preference by similarity to ideal solution (TOPSIS)

MCDA model, the TOPSIS [27], was applied for ranking algorithms [28] on different datasets. It computes the distances between a real solution and its ideal as well as negative ideal counterparts. We make use of the following algorithm

Step 1: Determine the normalized

$$r_{ij} = \frac{\chi_{ij}}{\sqrt{\sum_{j=1}^{J} \chi_{ij}^{2}}} \quad (9)$$

In formula (9), j = 1, 2..., J and I = 1, 2..., n represent the feature selection indices, respectively.
Step 2: Calculate weighted

$$v_{ij} = w_i r_{ij} \quad (10)$$

In formula (10), $w_i$ represents the weight
Step 3: Find the ideal solution $S^+$ and the negative ideal solution$S^-$

$$S^+ = \{v_1^+, , v_n^+\} = \left(\max v_{ij} | ij \in I'\right), (\min v_{ij} | i \in I'') \quad (11)$$

$$S^- = \{v_1^-, , v_n^-\} = \left(\min v_{ij} | i \in I'\right), (\max v_{ii} | i \in I'') \quad (12)$$

In formula (11) and (12), I′ and I″ represent the benefit criteria and cost criteria, respectively. Stability, TPR, TNR, accuracy and AUC are benefit criteria. Runway

Step 4: Calculate the Euclidean distance between the real and ideal solutions.

$$D_j^+ = \sqrt{\sum_{i=1}^{n} \left(v_{ij} - v_i^+\right)^2} \quad (13)$$

$$D_j^- = \sqrt{\sum_{i=1}^{n}\left(v_{ij} - v_i^-\right)^2} \quad (14)$$

Step 5: Calculate $R_j^+$.

$$R_j^+ = \frac{D_j^-}{D_j^+ + D_j^-}$$

Step 6: Rank feature selection methods by maximizing $R_j^+$.

| Data sets | Rank 1 | Rank 2 | Rank 3 | Rank 4 | Rank 5 | Rank 6 |
|---|---|---|---|---|---|---|
| **UIMS** | **GARF** | NN | MARS | SWR | SVM | CART |
| **QUES** | NN | **GARF** | MARS | SVM | CART | SWR |
| **CM1** | NN | MARS | **GARF** | CART | SWR | SVM |
| **JM1** | **GARF** | NN | SVM | MARS | CART | SWR |
| **KC1** | **GARF** | MARS | NN | CART | SVM | SWR |
| **KC3** | MARS | **GARF** | SWR | NN | SVM | CART |
| **MC1** | **GARF** | CART | MARS | SVM | NN | SWR |
| **MC2** | MARS | SWR | **GARF** | SVM | CART | NN |
| **MW1** | **GARF** | MARS | NN | SVM | SWR | CART |
| **PC1** | MARS | **GARF** | CART | SVM | NN | SWR |
| **PC2** | **GARF** | MARS | NN | CART | SVM | SWR |
| **PC3** | MARS | **GARF** | SVM | CART | SWR | NN |
| **PC4** | **GARF** | CART | MARS | SVM | SWR | NN |
| **PC5** | **GARF** | NN | CART | SVM | MARS | SWR |

**Table 3.** The overall analysis of the proposed model with other models.

The results of all the performance measures (MAE, RSQ, and RMSE) on various datasets are presented in Fig. 4 to Fig. 17. Based on UIMS, QUES, CM1, and JM1 data sets the best algorithm is NN for predicting the maintainability, (RMSE(UIMS)=8.57%, RMSE(QUES)=0.7%, RMSE(CM1) =1.46%, RMSE(JM1) =0.91%). The best algorithm for the datasets KC1, MW1, PC2, PC4, PC5, is GARF [29] with (RMSE(KC1) =0.13%, RMSE(MW1) =1.6%, RMSE(PC2) =0.5%, RMSE(PC4) =0.4%, RMSE(PC5) =0.2%). MARS algorithm is the best one for predicting the maintainability for the datasets KC3, MC2, PC1, and PC3, bearing the (RMSE(KC3) =1.2%, RMSE(MC2) =1.4%, RMSE(PC1) =1.0%, RMSE(PC3) =0.6%). Finally, the best algorithm for predicting the maintainability of the dataset MC1 is CART (RMSE (MC1) =0.77%). These results reveal that GARF is the best algorithm for 36% of the data sets, followed by NN, MARS (29%). The worst algorithms for predicting maintainability are CART (1%), stepwise, and SVM.

The results show that the best models in SMP are GARF, NN, and MARS. These SMP models outperformed other models on heterogeneous datasets. The GARF algorithm also outperformed other algorithms on datasets with a large number of instances and features. It is interesting to note that the performance of GARF has decreased with some datasets that have a large number of features but a small number of instances. It is understood that GARF works best when the search space is large and there are a large number of parameters involved.

In this study, the required data is collected from the public repositories, which were being used by several researchers in maintainability prediction. There are 12 NASA data sets (CM1, JM1, KC1, KC3, MC1, MC2, MW1, PC1, PC2, PC3, PC4, and PC5), and two popular private datasets (UIMS, QUES) that are taken up to evaluate the maintainability prediction on various algorithms GARF, NN, MARS, SWR, SVM, and CART. A popular multi-criteria decision-making technique [30], the TOPSIS method, was applied for ranking algorithms on different data sets. The ranking algorithms on different data sets are depicted in Table 3. From Table 3, it is observed that GARF is the best model for predicting the maintainability of automated software.

## 5. Threat to Validity

Limitations encountered during this study are listed below:

1. The results obtained in this study are based on NASA, Li & Henry datasets which were developed using C, C++, and Java. This study's model is applicable to those paradigms only. Further research on various languages can be conducted to improve usability.
2. The metrics considered for this study can be expanded with new metrics that influence software code and design. It can play a significant role in predicting maintainability.
3. The NASA datasets are from automated satellite applications, but the use of real-time advanced automated application data will improve the SPM's reliability.

4. The confidentiality of the datasets that will be shared for maintainability prediction remains a concern.
5. The GARF algorithm takes a huge amount of processing time and computational resource to produce the prediction results compared to other SMP models.

## 6. Conclusion

This paper focused on removing the ambiguity in maintainability prediction models for predicting the maintainability of heterogeneous automated applications. In this concern, various popular publicly available datasets of heterogeneous applications are considered and maintainability is predicted using six popular techniques, namely, step-wise regression, support vector machine, neural networks, multivariate adaptive regression splines, genetic algorithm and classification And Regression Tree. To choose the best model for predicting maintainability technique, multiple criteria decision-making model named TOPSIS is considered. The overall analysis has shown the efficiency of the proposed model over other popular maintainability prediction models. A range of possible future works have been identified while doing this research. There is a need of adapting real-time heterogeneous data for predicting maintainability. In addition, many other techniques and metrics should be further adapted to enhance the estimation of maintainability.

### Availability of Data and Materials

The datasets UIMS and QUES analyzed during the current study are available in the Li, Wei, and Sallie Henry. "Object-oriented metrics that predict maintainability." Journal of systems and software 23, no. 2 (1993): 111-122.
The datasets JM1, KC1, KC3, MC1, MC2, MW1, PC1, PC2, PC3, PC4, and PC5 analyzed during the current study are available in the [Promise] and [Mendeley Data] repositories,
[http://promise.site.uottawa.ca/SERepository/datasets-page.html]
[https://data.mendeley.com/datasets/923xvkk5mm/1]